\documentclass[twocolumn,
superscriptaddress,aps,10pt,longbibliography]{revtex4-1}

\usepackage[usenames,dvipsnames]{color}
\usepackage[colorlinks=true,citecolor=blue,linkcolor=magenta]{hyperref}
\usepackage{tikz-cd}
\usepackage{xcolor}
\usepackage{amsfonts,amssymb}
\usepackage[sumlimits,intlimits]{amsmath}
\usepackage{graphics}
\usepackage{graphicx}
\usepackage{mathrsfs}
\usepackage{textcomp}
\usepackage{verbatim}
\usepackage{bm}          

\newcommand{\be}{\begin{equation}}
\newcommand{\ee}{\end{equation}}

\newcommand{\bit}{\begin{itemize}}
\newcommand{\eit}{\end{itemize}}

\newcommand{\f}{\frac}
\renewcommand{\>}{\right\rangle}
\newcommand{\<}{\left\langle}
\newcommand{\ba}{\begin{align}}
\newcommand{\ea}{\end{align}}

\newcommand{\bi}{\begin{itemize}}
\newcommand{\ei}{\end{itemize}}
\newcommand{\lf}{\left(}
\newcommand{\ri}{\right)}
\newcommand{\dd}{\mathrm{d}}

\newcommand{\Tr}{\operatorname{Tr}}

\newcommand{\SO}{\mathrm{SO}}
\newcommand{\Og}{\mathrm{O}}
\newcommand{\SU}{\mathrm{SU}}

\graphicspath{{}{./images/}}

\begin{document}

\title{Note on Wess-Zumino-Witten models and quasiuniversality in 2+1 dimensions}

\author{Adam Nahum}
\affiliation{Theoretical Physics, University of Oxford, Parks Road, Oxford OX1 3PU, United Kingdom}

\date{\today}

\begin{abstract}
\noindent
We suggest the possibility that the two-dimensional $\SU(2)_k$ Wess-Zumino-Witten (WZW) theory, which has global $\SO(4)$ symmetry, can be continued to ${2+\epsilon}$ dimensions by enlarging the symmetry to ${\SO(4+\epsilon)}$. 
This is motivated by the three-dimensional sigma model with $\SO(5)$ symmetry and a WZW term, which is relevant to deconfined criticality.
If such a continuation exists, the structure of the renormalization group flows at small $\epsilon$ may be fixed by assuming analyticity in $\epsilon$.
This leads to the conjecture that the WZW fixed point annihilates with a new, unstable fixed point at a critical dimensionality $d_c>2$. 
We suggest that $d_c < 3$ for all $k$, and we compute $d_c$ in the limit of large $k$.
The flows support the conjecture that the deconfined phase transition in $\SU(2)$ magnets is a ``pseudocritical'' point with approximate $\SO(5)$, controlled by a fixed point slightly outside the physical parameter space.
\end{abstract}
\maketitle

This note  makes a conjecture about renormalization group (RG) flows in nonlinear sigma models (NL$\sigma$Ms) with WZW terms in $2+\epsilon$ dimensions. 
It is speculative, since we do not provide a concrete definition of these models in noninteger dimensions.
But we point out that assuming the existence of such a continuation in $\epsilon$ leads to interesting conclusions. 
The WZW fixed point survives up to a critical $\epsilon$, at which it annihilates with a new unstable fixed point that did not exist in 2D.
This critical $\epsilon_c$ can be  calculated easily only at large $k$, where $k$ is the WZW level, but we conjecture that for all $k$ the annihilation occurs in between 2 and 3 dimensions.
Our motivation is the case $\epsilon=1$, which is the $\SO(5)$--symmetric NL$\sigma$M for a 5-component unit vector, in 3D. This is a useful effective field theory for various interesting phase transitions \cite{tanakahu,tsmpaf06,deccp} that show numerical evidence of emergent $\SO(5)$ \cite{emergentso5,SandvikSpectrum,sreejith2018emergent,ippoliti2018half,li2019deconfined}.
The scenario obtained here supports, and gives a new way of thinking about, the ``quasiuniversal'' or ``pseudocritical'' RG flows conjectured previously  for these models \cite{DCPscalingviolations,wang2017deconfined}, 
since the fixed point annihilation at $d_c \lesssim 3$ suggested by this calculation provides a mechanism for slow RG flows in $d=3$. We return to this at the end.

The Euclidean action for the $\SU(2)_k$ WZW model in 2D,  in terms of an $\SU(2)$ matrix $g(x_1, x_2)$, is \cite{witten1984non,polyakov1983theory,knizhnik1984current,affleck1987critical,francesco2012conformal}
\be
S = \f{1}{2\lambda^2} \int \dd^2 x \, \Tr \, (\partial_\mu g^{-1}) (\partial_\mu g) + i k\, \Gamma.
\ee
$\Gamma$ is the WZW term, written in terms of an  extension $g(x_1, x_2, x_3)$ of the field to a fictitious 3D ``bulk''  as ${\Gamma = \f{\epsilon_{\mu\nu\lambda} }{12\pi} \int \dd^3 x \Tr (g^{-1}\partial_\mu g) (g^{-1}\partial_\nu g) (g^{-1}\partial_\lambda g)}$. 
The field lives on the sphere $S^3$, and can be written as a four-component unit vector $\Phi$ using the Pauli matrices: ${g = \Phi_0 \mathbb{I} + i \sum_{a=1}^3 \Phi_a \sigma^a}$. 
Therefore this is also the standard $\Og(4)$ sigma model, with the addition of the WZW term, which reduces the internal symmetry to ${\SO(4)=[{\SU(2)_\text{L}\times \SU(2)_\text{R}}]/{\mathbb{Z}_2}}$.
For a given $k\in \mathbb{Z}$, the theory has an unstable, trivial fixed point at $\lambda^2=0$, and a stable, nontrivial one at ${\lambda_*^2={4\pi}/{|k|}}$ \cite{witten1984non,francesco2012conformal}. 

The construction generalizes to $d$ dimensions, giving the NL$\sigma$M for a $(d+2)$-component ``spin'', with a WZW term and $\SO(d+2)$ symmetry (see e.g. \cite{abanov2000theta}):
\be
S_d = \f{1}{\lambda^2} \hspace{-0.5mm}
\int 
\hspace{-0.5mm} (\partial \Phi)^2 
+ 
\f{2\pi i k \epsilon_{a_1\ldots a_{d+2}}}{\text{area}(S^{d+1})} 
\hspace{-0.5mm}
\int 
\hspace{-0.5mm}
\Phi_{a_1} \partial_{x_1} \Phi_{a_2} \ldots \partial_{u} \Phi_{a_{d+2}}
\ee
The most interesting case for us in the above hierarchy of theories is $S_3$, the $\SO(5)$ sigma model in $d=3$. 
In $d=1$ the standard kinetic term is irrelevant at low energies, and dropping it leaves the usual coherent-states path integral for a spin of size $k/2$ \cite{wiegmann1988superconductivity}. 
The $d=0$ case is an integral: writing ${\Phi_0+i\Phi_1=e^{i\theta}}$, the action is $S_0=i k \theta$, and the ``correlator'' is $\< e^{im\theta}\>= \delta_{m,k}$.

These theories, often with symmetry-breaking anisotropy terms,
have many applications to critical phenomena. These applications can usually be understood heuristically  from the fact that $S_\ell$ is the effective theory on an appropriate $\ell$-dimensional defect
(built by fixing the configuration of $d-\ell$ components of $\Phi$)
 in the $d$-dimensional theory $S_d$.
For example, we may construct a hedgehog-like configuration for $d$ components of $\Phi$. The effective theory at this defect is $S_0$ for the remaining two components. The above expression for $\<e^{im\theta}\>$ then shows that such defects are forbidden except at the loci of insertions of $e^{i\theta(x)}$. This is connected to the fact that an anisotropic version of $S_3$ describes the 3D $\Og(3)$ model with hedgehog defects forbidden \cite{lau1989numerical,kamal1993new,motrunichvishwanath1,tsmpaf06,sreejith2018emergent}.

Motivated by this hierarchy of field theories, let us entertain the possibility that the fixed points present in 2D can be tracked to $2+\epsilon$ dimensions. 
Whether this can be made precise is less clear than in the case without a WZW term, where the $2+\epsilon$ expansion is standard, because the structure of the topological term depends on the dimensionality \footnote{A framework for dimensional regularization of the 2D WZW model has been developed \cite{bos1987dimensional,bos1988example,xi1988three,zheng1989dimensional,de1993wznw,ali2001four}. However it does not retain Lorentz invariance in $d\neq 2$, so is not suitable for our purpose here.}. 
Nevertheless, if we assume the continuation exists, the flows at small $\epsilon$ can be fixed very simply using known results in 2D and assuming analyticity of the RG equations in $\epsilon$.
This is inspired by the treatment of the $\Og(n)$ model close to $n=d=2$  in Ref.~\cite{cardy1980n}.

\begin{figure}[t]
\centering
  \includegraphics[width=0.8\columnwidth]{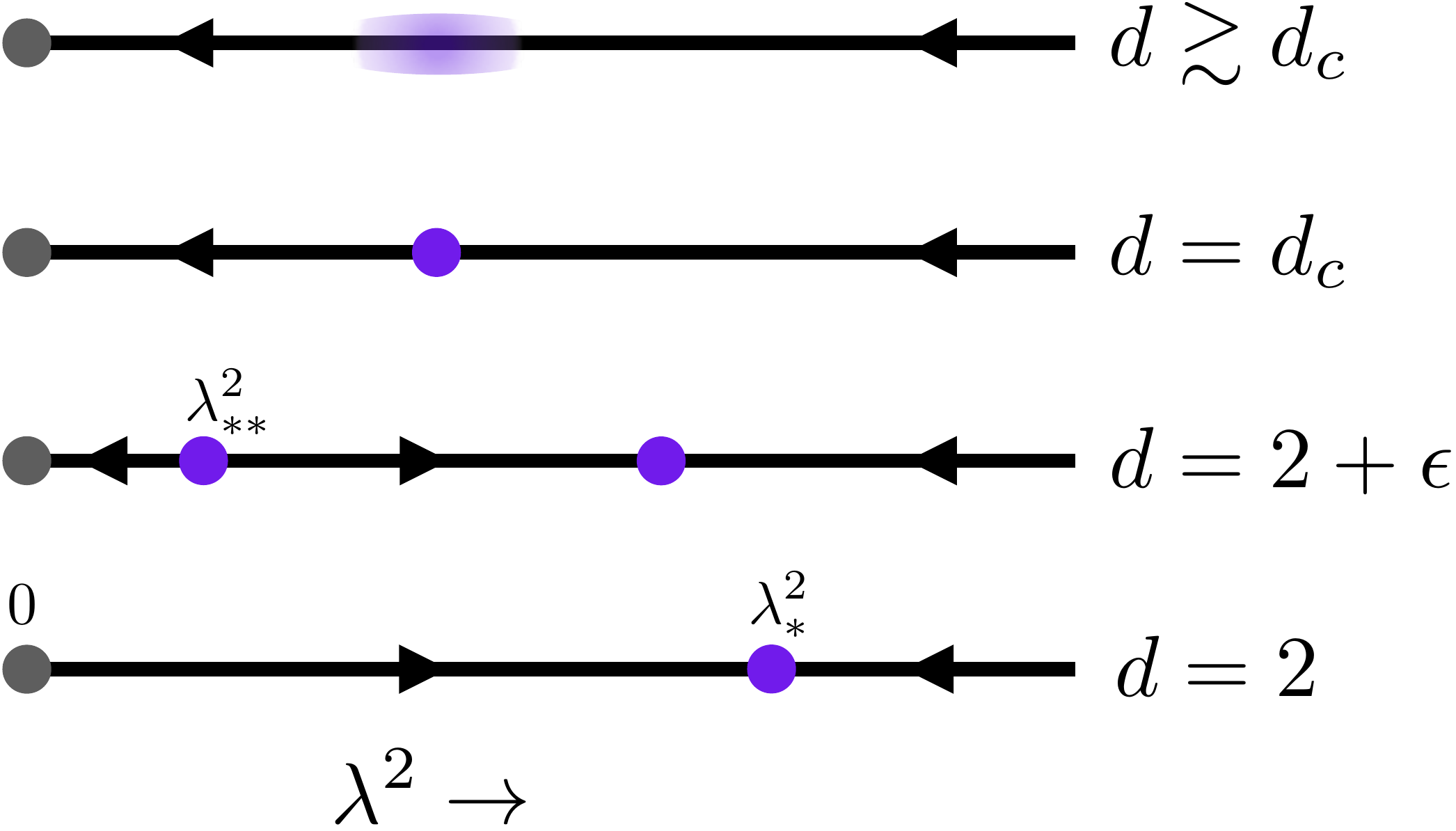}
\caption{Topology of flows, as a function of dimension. The smudge ($d\gtrsim d_c$) indicates slow RG flow, without a fixed point.}
\label{fig:flowtopo}
\end{figure}

In two dimensions the one-loop beta function is \cite{witten1984non}
\be\label{eq:1loop}
\f{\dd \lambda^2}{\dd \ln L}
 = 
 \f{\lambda^4}{2\pi}
  \lf
   1 - \lf \f{\lambda^2 k}{4 \pi } \ri^2
    \ri
\ee
The one-loop approximation is justified at large $|k|$ because the fixed point is at $\lambda^2 = \mathcal{O}(k^{-1})$, so that the entire action is multiplied by a large parameter of order $k$ \cite{witten1984non}.
For $k$ of order 1 we should use an unknown exact $\beta$ function,  but with the same topology of flows. We write this schematically as
\be
\f{\dd \lambda^2}{\dd \ln L}
 = 
\beta_k^{(0)}(\lambda^2).
\ee
We now go to $d=2+\epsilon$, assuming the RG equations are analytic in $\epsilon$:
\be
\f{\dd \lambda^2}{\dd \ln L}
 = 
\beta_k^{(0)}(\lambda^2) + \epsilon \, \beta_k^{(1)} (\lambda^2) + \mathcal{O}(\epsilon)^2.
\ee
In the limit of small $\lambda^2$ we have, trivially,
\ba
 \beta_k^{(0)} (\lambda^2) & = \f{\lambda^4}{2\pi} + \mathcal{O}(\lambda^6),
 &
 \beta_k^{(1)} (\lambda^2) & = - \lambda^2 +  \mathcal{O}(\lambda^4). 
\end{align}
This is already enough to fix the topology of the RG flows when $\epsilon$ is small: see Fig.~\ref{fig:flowtopo}, third panel.
At $\epsilon=0$ we have a marginally unstable fixed point at $\lambda^2=0$ and a stable one at $\lambda_*^2$. 
The latter remains stable and isolated for small $\epsilon$ (but, if the signs predicted by the perturbative expressions are valid, it shifts towards the origin by $\mathcal{O}(\epsilon)$, and its irrelevant RG eigenvalue moves slightly towards zero).
In contrast, the perturbation splits the fixed point at $\lambda^2=0$ into a stable fixed point at $\lambda^2=0$ and an \textit{unstable} fixed point at $\lambda^2_{**} \simeq 2 \pi \epsilon$.
This splitting in the vicinity of $\lambda^2=0$ is similar to the $\Og(N)$ NL$\sigma$M without a WZW term; in both cases the unstable fixed point governs a transition between phases with broken/unbroken symmetry. Here however the universality class of the fixed point at $\lambda^2_{**}$  is different, as is that of the unbroken phase.

The likely situation is that, at some $d_c(k)$, the unstable fixed point which is moving away from the origin collides and annihilates with the stable fixed point which is moving towards the origin --- so that in high dimensions there is no fixed point for real $\lambda^2$. At $d=d_c(k)$ we have a marginally stable fixed  point (Fig.~\ref{fig:flowtopo}).

\begin{figure}[t]
\centering
  \includegraphics[width=0.75\columnwidth]{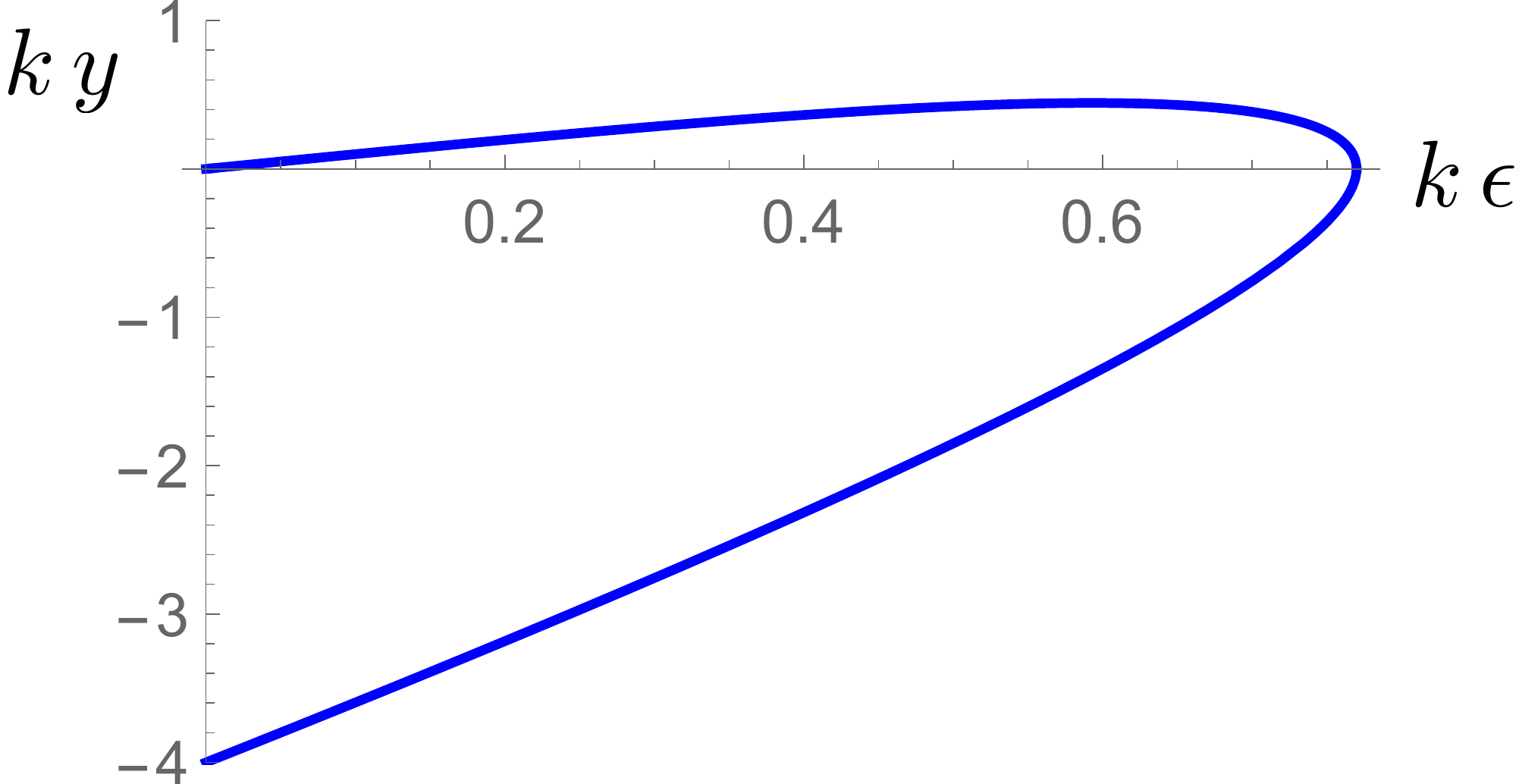}
\caption{RG eigenvalues $y$ for stable (lower branch) and unstable (upper) fixed points as a function of $
\epsilon$ at large $k$.}
\label{fig:rgeigs}
\end{figure}

We can be more concrete when  $k$ is large. Consider the scaling $k\gg 1$ with $\epsilon k$ of order 1. The relevant regime is where $\lambda^2$ is of order $\epsilon$. The leading terms are then:
\be\label{eq:1loopepsilon}
\f{\dd \lambda^2}{\dd \ln L}
 = - \epsilon \lambda^2 +
 \f{\lambda^4}{2\pi}
  \lf
   1 - \lf \f{\lambda^2 k}{4 \pi } \ri^2
    \ri.
\ee
We see that the annihilation described above indeed occurs, and the critical dimensionality is:
\be
d_c(k) = 2+ \f{4}{3 \sqrt{3} \times k}.
\ee
Fig.~\ref{fig:rgeigs} shows the RG eigenvalues of the stable and unstable fixed points for $d<d_c$.

When $d \gtrsim  d_c$ we have pseudocritical RG flows. 
Slow flow for $\lambda^2 \sim \f{4\pi}{\sqrt{3} k}$, where the flows are approximately 
\be
\f{\dd \delta \lambda^2}{\dd \ln L} \simeq - \f{4\pi (d-d_c)}{\sqrt{3} k} - \f{(\delta \lambda^2)^2}{2\pi},
\ee
yields the exponentially large correlation length  ${\xi \sim \exp \f{3^{1/4}\pi \sqrt{k}}{\sqrt{2(d-d_c)}}}$,
as in other theories with a fixed point annihilation  
\cite{nienhuispotts,cardynauenbergscalapino,Gies,kaplan,PhaseTransitionsCPNSigmaModel,DCPscalingviolations,giombi2016conformal,herbut2016chiral,gukov2017rg}.
Ref.~\cite{wang2017deconfined} argued that in such a situation,  expanding the RG equations for irrelevant couplings in ${d-d_c}$ shows that quasiuniversality (independence of UV couplings) holds on long scales, 
to exponentially good precision in $[d-d_c]^{-1/2}$, despite the fact that $\lambda^2$ drifts: different microscopic models travel along the same quasiuniversal flow line in theory space. For $d\gtrsim d_c$ we also have complex, $\SO(d+2)$-symmetric fixed points with $\operatorname{Im}\lambda^2\propto \sqrt{d-d_c}$.
Complex fixed points have been explored recently in Refs.~\cite{gorbenko2018walking,gorbenko2018walking2,ma2019shadow,benvenuti2018qed,faedo2019holographic}.

In the context of deconfined criticality we are interested in 3D models that in the UV have a smaller symmetry than $\SO(5)$. 
If $d_c$ is close enough to 3 to give a large $\xi$ in 3D,
and assuming that the four-index symmetric tensor of  $\SO(4+\epsilon)$  is irrelevant at $d_c$ \cite{emergentso5,wang2017deconfined} (this is the case at large $k$, where scaling dimensions are close to those in 2D)
then the above flows will lead to a pseudocritical phase transition with approximate emergent $\SO(5)$, by the scenario discussed in Refs.~\cite{wang2017deconfined,serna2019emergence}. 
This scenario is consistent with simulations, and, since it does not require a unitary 3D fixed point, with conformal bootstrap \cite{bootstrappingO(N),SimmonsDuffinSO(5),Nakayama,poland2019conformal}.
It is also consistent with what we know about various dual gauge theories for deconfined criticality \cite{wang2017deconfined,DCPscalingviolations}, including recent $\epsilon$-expansion results \cite{ihrig2019abelian,janssen2017critical, ihrig2018deconfined}.
The endpoint of the quasiuniversal flow line is the ordered phase ($\lambda^2=0$): in the application to deconfined criticality this means that at the very longest scales the emergent symmetry gets spontaneously broken, giving artificial ${\SO(5)}$ ``Goldstone modes'' with a very small mass \cite{zhao2019symmetry,serna2019emergence}.

Though speculative, the present lowest-order expansion supports this scenario. If a consistent framework for expanding to higher orders in $\epsilon$ \footnote{At large $k$ both nontrivial fixed points (stable and unstable) are close to the origin.
If we fix $k$ and treat only $\epsilon$ as small, then it is the unstable fixed point that is accessible.}  can be defined,
then this would be one way to  put the pseudocriticality scenario for $\SO(5)$ on firm ground.
The above also suggests examining numerically the 3D  models with $k>1$ (or rather related sign-free lattice models  which could be based on those relevant to the $k=1$ case \cite{SandvikJQ, DCPscalingviolations,ippoliti2018half,sreejith2018emergent}), to test for pseuducriticality there.

We can consider other, related deformations of the WZW model.  At the order to which we have worked,  changing the dimension to $2+\epsilon$ has the same effect on the RG flows as changing  the power of momentum $q$ in the kinetic term to $|q|^{2-\epsilon}$. 
This raises the question of whether we can study quasiuniversality in the 3D model, while avoiding the WZW term in noninteger dimensions,  by imposing a dispersion of the form $|q|^{3-\delta}$ with $\delta>0$.
It also raises the question of whether we can obtain pseudocriticality, fixed point annihilation, complex fixed points, etc.,  in the \textit{one}-dimensional (0+1D)  model with a WZ term, by taking a coupling that  is long-ranged \cite{kosterlitz1976phase,brezin1976critical}
 in time, $\sim |t-t'|^{-(2-\delta)}$, and varying $\delta$.
This model is relevant to the dynamics of a spin coupled to a bath \cite{anderson1970exact,sachdev2004quantum,vojta2006impurity}. We hope to return to these issues elsewhere.

 \textit{Related work:}
After completion of this work, I became aware of independent work by Ruochen Ma and Chong Wang reaching the same essential conclusions (to appear in the same arXiv posting). 

\textit{Acknowledgments:} I thank John Chalker, Patrick Draper, Fabian Essler,
John March-Russell, Michael Scherer, and T. Senthil for useful discussions. I thank Ruochen Ma and Chong Wang for sharing results prior to publication. This work was supported by a Royal Society University Research Fellowship.

\bibliographystyle{unsrt} 
\bibliography{WZWrefs}

\end{document}